\documentclass[journal=jacsat,manuscript=article]{achemso}
\usepackage{orcidlink}

\usepackage[version=3]{mhchem} 

\author{Shreyas S. Joglekar \orcidlink{0000-0001-9354-2666}}
\altaffiliation{Equal contribution}
\author{Korbinian Baumgaertl \orcidlink{0000-0002-9410-1931}}
\altaffiliation{Equal contribution}
\author{Andrea Mucchietto \orcidlink{0000-0001-5562-3916}}
\author{Francis Berger}
\author{Dirk Grundler \orcidlink{0000-0002-4966-9712}}
\email{dirk.grundler@epfl.ch}
\affiliation[Ecole Polytechnique Fédérale de Lausanne (EPFL)]
{Laboratory of Nanoscale Magnetic Materials and Magnonics, Institute of Materials (IMX), Ecole Polytechnique Fédérale de Lausanne (EPFL), 1015 Lausanne, Switzerland}

\alsoaffiliation[Second University]
{Institute of Electrical and Micro Engineering (IEM), EPFL, 1015 Lausanne, Switzerland}

\title
  {Magnetization Reversal of 50-nm-wide Ni$_{81}$Fe$_{19}$ Nanostripes by Ultrashort Magnons in Yttrium Iron Garnet for Memory-Enhanced Magnonic Circuits}

\begin{document}

\newpage 

\begin{abstract}
 
  Spin waves (magnons) can enable wave-based neuromorphic computing by which one aims at overcoming limitations inherent to conventional electronics and the von Neumann architecture. In this study, we explore the storage of magnon signals and the magnetization switching of periodic and aperiodic arrays of Ni$_{81}$Fe$_{19}$ (Py) nanostripes with widths $w$ between 50 nm and 200 nm. Spin waves excited with low microwave power in yttrium iron garnet induce the reversal of the nanostripes of different $w$ in a small opposing field. Exploiting microwave-to-magnon transducers for magnon modes with ultrashort wavelengths $\lambda$, we demonstrate the reversal of 50-nm-wide Py nanostripes by magnons with $\lambda\approx 100~$nm after they have propagated over 25 $\mu$m in YIG. The findings are important for designing a magnon-based in-memory computing device.  
\end{abstract}

\section{Keywords}
magnons, ferromagnet, ferrimagnet, magnetization reversal, microwaves 
\newpage
\section{Introduction}
The use of various Artificial Intelligence platforms like Dall-E2 \cite{Dalle}, ChatGPT \cite{ChatGPT} has skyrocketed in recent months. Such platforms use Machine Learning algorithms to generate a tremendous amount of data in computers' processors and store it in separate memory units leading to enormous data trafficking, reduced computing performance and Joule heating losses. These drawbacks of the conventional technology motivate the search for alternatives such as in-memory and neuromorphic computing \cite{Inmem_Sebastian2020}. These alternatives require two essential components: nonlinearity in signal processing and nonvolatile memory. 

Recently, magnonic neural networks have been proposed which exploit wave-based signal processing in a ferrimagnetic thin film combined with an array of nanostructured ferromagnets. Exploiting propagating spin waves (magnons) in yttrium iron garnet (YIG) and their interference underneath nanomagnets, neural network functionality was demonstrated in that the magnetic states of bistable nanomagnets were iteratively modified and programmed \cite{MagnonicsRoadmap2022,csaba,wang2023perspective}. 
The spin waves are collective excitations of spins that transfer angular momentum in a magnetically ordered material. In a low-damping material like YIG they can propagate over long distances up to the mm length scale \cite{Maendl2017}. They possess engineered wavelengths down to a few tens of nanometers when excited on-chip by microwaves at GHz frequencies \cite{kbacs}. The interaction with distributed magnets controls their scattering, phases and interference \cite{csaba,HYu2022}. A similar hybrid system gives rise to the magnonic holographic memory (MHM) in which coherent spin waves read out stored data via interference over several unit cells (magnetic bits) \cite{khitun2023}. However, for both the magnonic neural network and MHM, it is not yet clear how to reprogram the magnets and encode the data. Beyond applying global magnetic \cite{Lianze2022} or electrical fields \cite{Zhu2017}, local laser heating \cite{C9NR01628G} or the tip of a magnetic force microscope \cite{Gartside2018} might be used to reverse nanomagnets, but magnetization reversal induced by spin waves avoids additional equipment and infrastructure.

Magnetization switching attributed to magnons in antiferromagnetic NiO was reported by Wang et al. \cite{Switch2019} and Guo et al. \cite{switch2021}. The type and wavelengths of spin waves leading to magnet reversal were not explored however. Baumgaertl et al. \cite{Baumgaertl2023} demonstrated that spin waves propagating in YIG induced magnetization reversal of 100 nm wide and 25 to 27 $\mu$m long Permalloy (Ni$_{81}$Fe$_{19}$ or Py) nanostripes in a small opposing field $H$. They were separated from the spin-wave emitting coplanar waveguide (CPW) by a distance of $\geq 25~\mu$m. The relevant mode had a wavelength $\lambda$ of about 7 $\mu$m. It is now timely to investigate the magnon-induced reversal of nanostripes of different widths and spin waves of shorter wavelengths.

In this work we report magnon-induced reversal by propagating spin waves with $\lambda$ down to 100 nm. Nanostripe widths are varied from 50 to 200 nm. Going beyond the earlier experiments, we explore the reversal in periodic and aperiodic arrays of nanostripes of different widths, thereby avoiding a commensurability effect between $\lambda$ and period $a$. Spin waves with $\lambda$ close to 100 nm reverse nanostripes at a precessional power $P_{\rm prec}$ of 10.8 nW in a small opposing field. The value is of the same order as the one found for long-wavelength spin waves of $\lambda=7~\mu$m explored under similar conditions in Ref. \cite{Baumgaertl2023}. Our findings are promising for enhanced functionalities of magnonic neural networks and MHMs operating on the nanoscale.

\section{Results and discussion}

\subsection{Magnon mode dependent switching of 50-nm-wide nanostripes}

\begin{figure}[H]
    \centering
    \includegraphics[width = \linewidth]{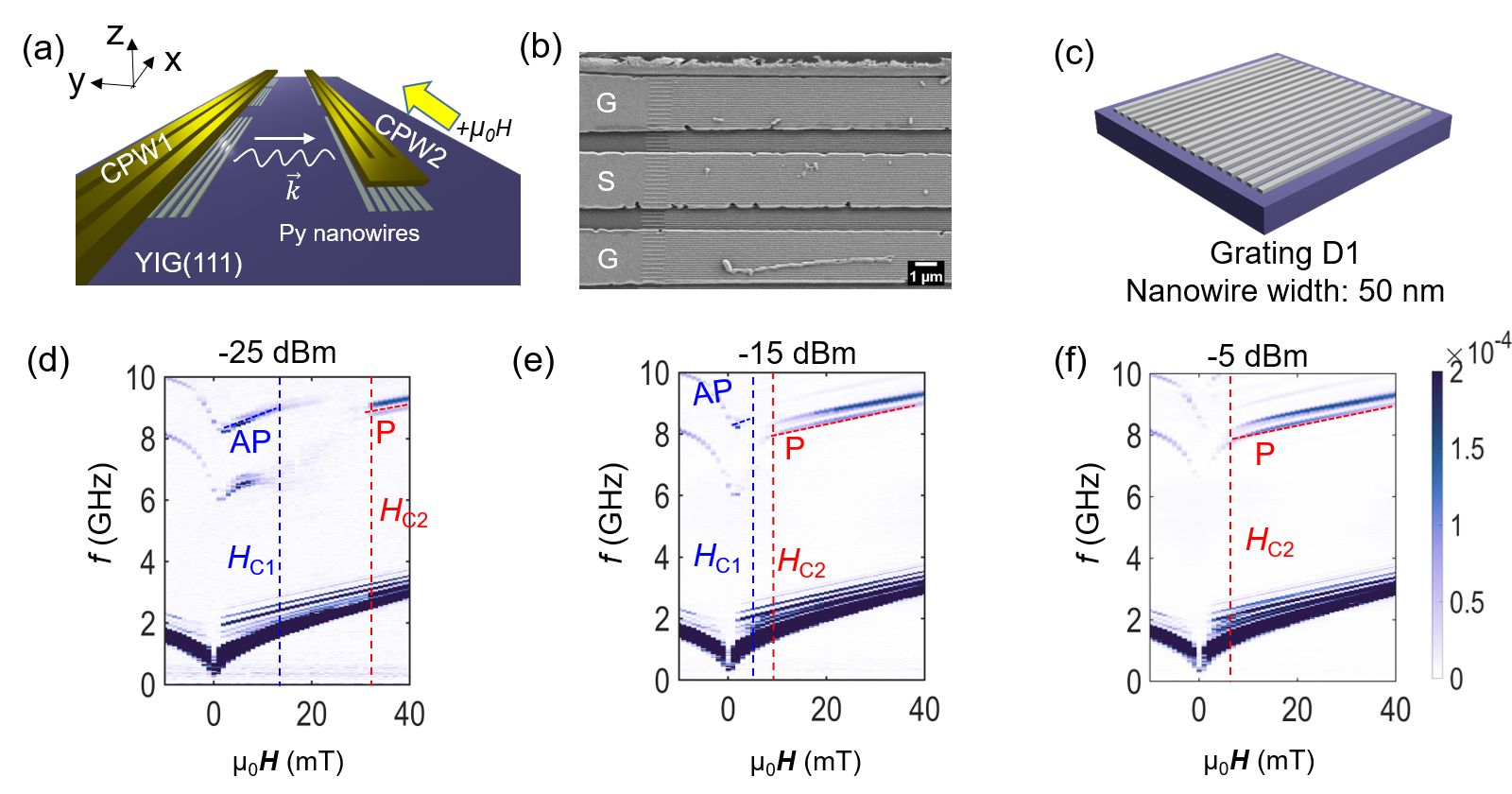}
    \caption{(a) Sketch of the experiment involving Py nanostripes fabricated on 100 nm-thin YIG (111) beneath CPWs separated by a signal-to-signal line distance of 35 $\mu$m. (b) Scanning Electron Micrograph of sample D1 showing the ground (G) and signal (S) lines of a CPW covering the Py nanostripes and the YIG film. The lengths of nanowires alternated between 25 and 27 $\mu$m consistent with Ref. \cite{Baumgaertl2023}. (c) Illustration of one of the nanostripe arrays (gratings) used in sample D1 with 50-nm-wide stripes and period $a=100~$nm. Scattering parameter Magnitude of $\Delta S21$ measured in transmission configuration between CPW1 and CPW2 using (d) $P_{\rm irr}=-25~$dBm, (e) $-15~$dBm and (f) $-5~$dBm. We extract fields $H_{\rm C1}$ and $H_{\rm C2}$ (indicated by vertical dashed lines) denoting the power-dependent onset and completion, respectively, of the nanostripes' reversal. 
    }
    \label{fig:fig1}
\end{figure}

Figure \ref{fig:fig1}(a) illustrates the experiment. Arrays of Py nanostripes (gratings) were fabricated on 100-nm-thick YIG grown on a Gadolinium Gallium Garnet GGG (111) substrate. In periodic gratings the 20-nm-thick stripes exhibited widths $w$ of either 50, 100 or 200 nm. The periods amounted to $a=2w$. The sample D1 with stripes of width $w=50~$nm is shown in Fig. \ref{fig:fig1} (b) and  sketched in (c). The period $a$ was 100 nm. In addition, we prepared and investigated a sample containing an aperiodic grating. Stimulated by Ref. \cite{Lisiecki2019}, we arranged nanostripes on a Fibonacci sequence. Thereby the stripe positions were not commensurate with a propagating magnon wavelength. CPWs were fabricated above the nanostripes at a signal-to-signal line (edge-to-edge) distance of 35 $\mu$m (25 $\mu$m). Spin waves were excited at the emitter CPW (CPW1) with port 1 of a Vector Network Analyzer (VNA) and their propagation was analyzed at the detector CPW (CPW 2) using VNA port 2. An in-plane magnetic field $\mu_0$\textbf{\textit{H}} was applied using current-controlled solenoids. 

We applied the following procedures to perform broadband spin-wave spectroscopy at different VNA powers $P_{\rm irr}$ between -30 and +15 dBm. The nanostripes were saturated in $-x$ direction using $\mu_0 H= -90$~mT. Spin waves were excited at the given $P_{\rm irr}$ for frequencies $f_{\rm irr}$ ranging from 10 MHz to 20 GHz with a step of 2.5 MHz. The magnetic field was increased to +40 mT in steps of 1 mT. Transmission spectra $S21$ were recorded at each value of $\mu_0 H$. The median value of the $S21$ spectra measured over the range of magnetic field was subtracted from the raw data to remove the background signal. Figure \ref{fig:fig1}(d-f) show the corresponding spectra $\Delta S21$ for three different values $P_{\rm irr}$ as indicated in the panels. The dark branches correspond to different spin wave modes propagating from CPW1 to CPW2. With increasing power, the high-frequency branches (grating-coupler modes) labelled by P start at smaller and smaller field $H$ (red dashed lines). At the same time, the branches AP end at smaller and smaller $H$ (blue dashed lines) or do not appear in panel (f). We attribute the P (AP) branches to a magnetic configuration in which Py nanostripes are aligned with (are anti-parallel to) the magnetization of the YIG. YIG reverses its magnetization at a field below 2 mT. In an intermediate field regime the high frequency branches AP and P are not resolved indicating that the gratings are not uniformly magnetized. This field regime reflects the switching field distribution of Py nanostripes. Following Ref.~\cite{Baumgaertl2023}, the disappearance and reappearance of high-frequency branches at positive fields define critical field values $H_{\rm C1}$ and $H_{\rm C2}$, respectively, which quantify the switching field distribution of the nanostripe arrays. The critical field $H_{\rm C1}$ ($H_{\rm C2}$) is given by the value of applied magnetic field $H$ at which the signal strength of the branch AP (P) has decreased (increased) to 50\% of its maximum signal strength. The two critical fields are attributed to the onset and completion of nanostripes' reversal. In sample D1, we extracted $H_{\rm C1}$ and $H_{\rm C2}$ considering the spin wave mode with wave vector $k=2\pi/\lambda=k_1+1G^{\rm D1}$ and $G^{\rm D1}=2\pi/a=2\pi/(100~{\rm nm})$. $k_1$ corresponds to the most prominent wave vector provided by the CPW and $G^{\rm D1}$ to the reciprocal lattice vector of sample D1. With increasing $P_{\rm irr}$, $H_{\rm C1}$ and $H_{\rm C2}$ decrease in Fig. \ref{fig:fig1}(d-f). In Fig. \ref{fig:fig1}(f), branch AP is no longer observed. This observation indicates that the reversal of 50-nm-wide nanostripes starts at an applied field as low as 1 mT for $P_{\rm irr}=-5~$dBm.  

\begin{figure}[H]
    \centering
    \includegraphics[width = \linewidth]{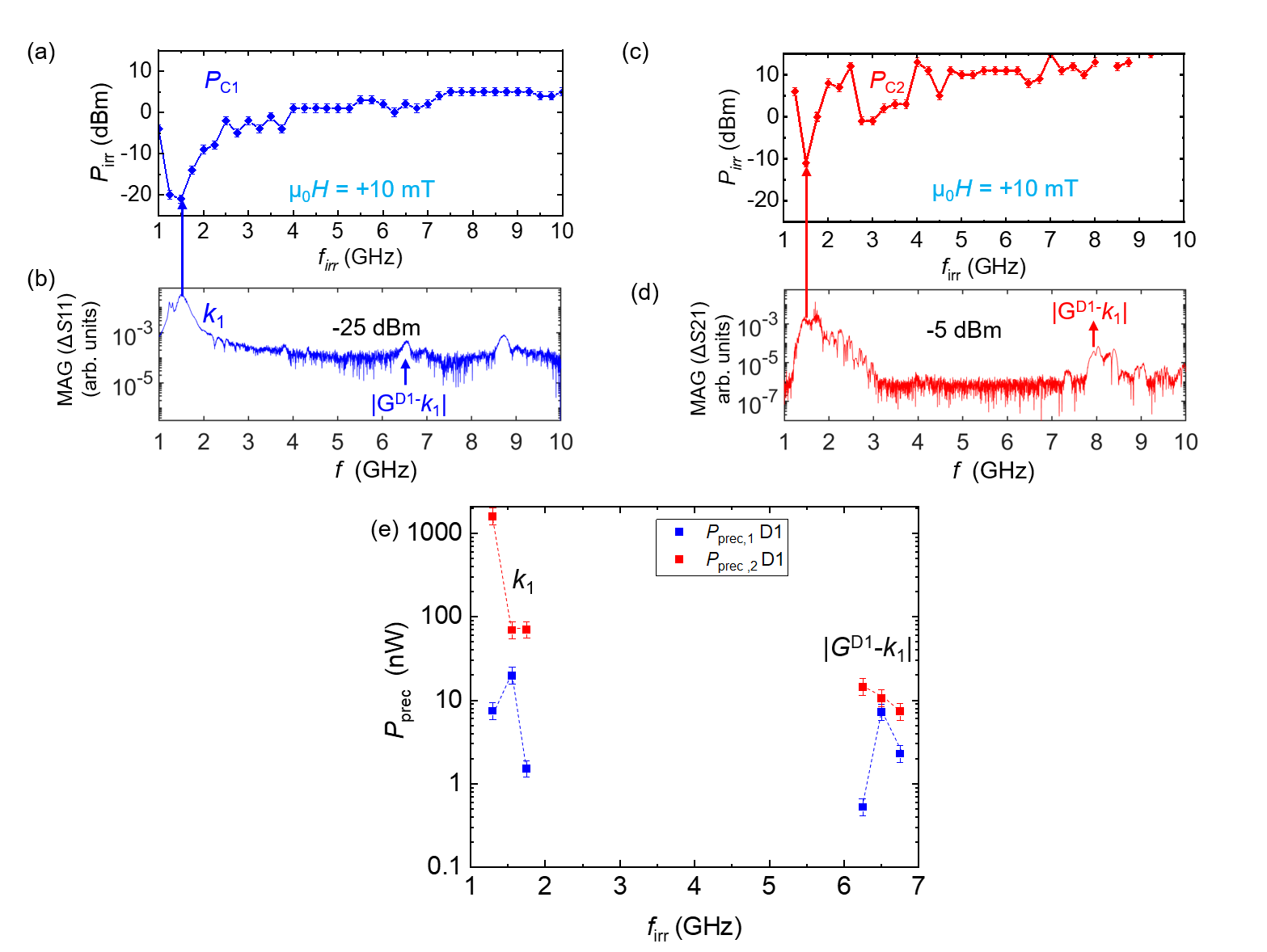}
    \caption{(a) Switching yield diagram of D1. The symbols interconnected by lines indicate the frequency-dependent critical power attributed to the reversal of about 50\% of the nanostripes below CPW1 at +10 mT. (b) Magnitude of $\Delta S11$ obtained at +10 mT with a low power $P_{\rm sens}$ of -25~dBm. (c) Switching yield diagram reflecting nanostripes' reversal below CPW2. (d) Magnitude of $\Delta S21$ measured at +10 mT using $P_{\rm sens}$ of -5~dBm. The spectrum differs from (b). A clear shift of mode $|G^{\rm D1}-k_1|$ from 6.5 GHz in (b) to 8 GHz in (d)  is resolved. For (d), we assume nanostripes beneath both CPWs to be switched by mode $k_1$ which is excited at a power larger than $P_{\rm C1}$ and $P_{\rm C2}$ at 1.5 GHz. (e) Precessional power values $P_{\rm prec}$ at frequencies corresponding to modes $k_1$ and $|G^{\rm D1}-k_1|$.}
    \label{fig:fig2}
\end{figure}
Before discussing the power dependence of critical fields in different samples it is instructive to identify the magnon modes in sample D1 which induce most efficiently nanostripe reversal in a fixed opposing field of +10 mT. This field value is smaller than $\mu_0H_{\rm C1}$ determined at -30 dBm. For different $f_{\rm irr}$ the power $P_{\rm irr}$ was increased until the branch AP (P) in $\Delta S21$ reduced (increased) to 50\% of its maximum signal strength (see methods and supporting information S1). These critical power values $P_{\rm C1}$ ($P_{\rm C2}$) are summarized in the switching yield diagrams of Fig. \ref{fig:fig2}(a) [(c)]. We find that the mode $k_{1}$ ($\lambda$ = 7.2 $\mu$m) which is directly excited by CPW1 at 1.5 GHz [Fig. \ref{fig:fig2}(b)] induces reversal of 50-nm-wide Py stripes at a power $P_{\rm C1}=P_{\rm irr}= -20$ dBm underneath CPW1. At $P_{\rm C2}=-11~$dBm, mode $k_1$ [Fig. \ref{fig:fig2}(d)] reversed about 50\% of nanostripes underneath CPW2. These observations are qualitatively consistent with Ref. \cite{Baumgaertl2023}. For the grating coupler mode with $k=|G^{\rm D1}-k_1|$ ($\lambda$ = 101.4 nm) the corresponding power values read $P_{\rm C1}=+2$ dBm and $P_{\rm C2}=+8$ dBm, respectively. We note that this magnon mode with ultrashort wavelength propagated over 25 $\mu$m through bare YIG and induced stripe reversal underneath CPW2. This was not resolved in Ref. \cite{Baumgaertl2023}.

To compare the efficiency in magnon-induced stripe reversal between different samples, we evaluate the power $P_{\rm prec}$ transferred to the spin precession (prec) in YIG \cite{Baumgaertl2023}. This parameter is independent of the individual microwave-to-magnon transduction of CPWs. The precessional power values $P_{\rm prec}$ are evaluated by multiplying the critical power $P_{\rm C1}$ or $P_{\rm C2}$ obtained from the switching yield diagrams with the square of the magnitude of $\Delta S11$ (see supporting information S2) measured at -30 dBm and -10 dBm, respectively. Figure \ref{fig:fig2}(e) summarizes precessional power values $P_{\rm prec,1}$ and $P_{\rm prec,2}$, respectively, for frequencies at which the magnon modes $k_1$ and $k=|G^{\rm D1}-k_1|$ are excited. We observe that mode $k_1$ provides $P_{\rm prec,2}$ of 68.9 nW. This value is similar to  the one found in Ref. \cite{Baumgaertl2023}. Strikingly, the grating coupler mode $|G^{\rm D1}-k_1|$ with an ultrashort wavelength of about 100 nm requires $P_{\rm prec,2}$ of only 10.6 nW for reversing stripes. The data demonstrate that short-wavelength magnon modes are more efficient than long-wavelength modes in terms of precessional power needed for reversal. 

\subsection{Magnon-induced switching of nanostripes in an aperiodic grating}

\begin{figure}[H]
    \centering
    \includegraphics[width= \linewidth]{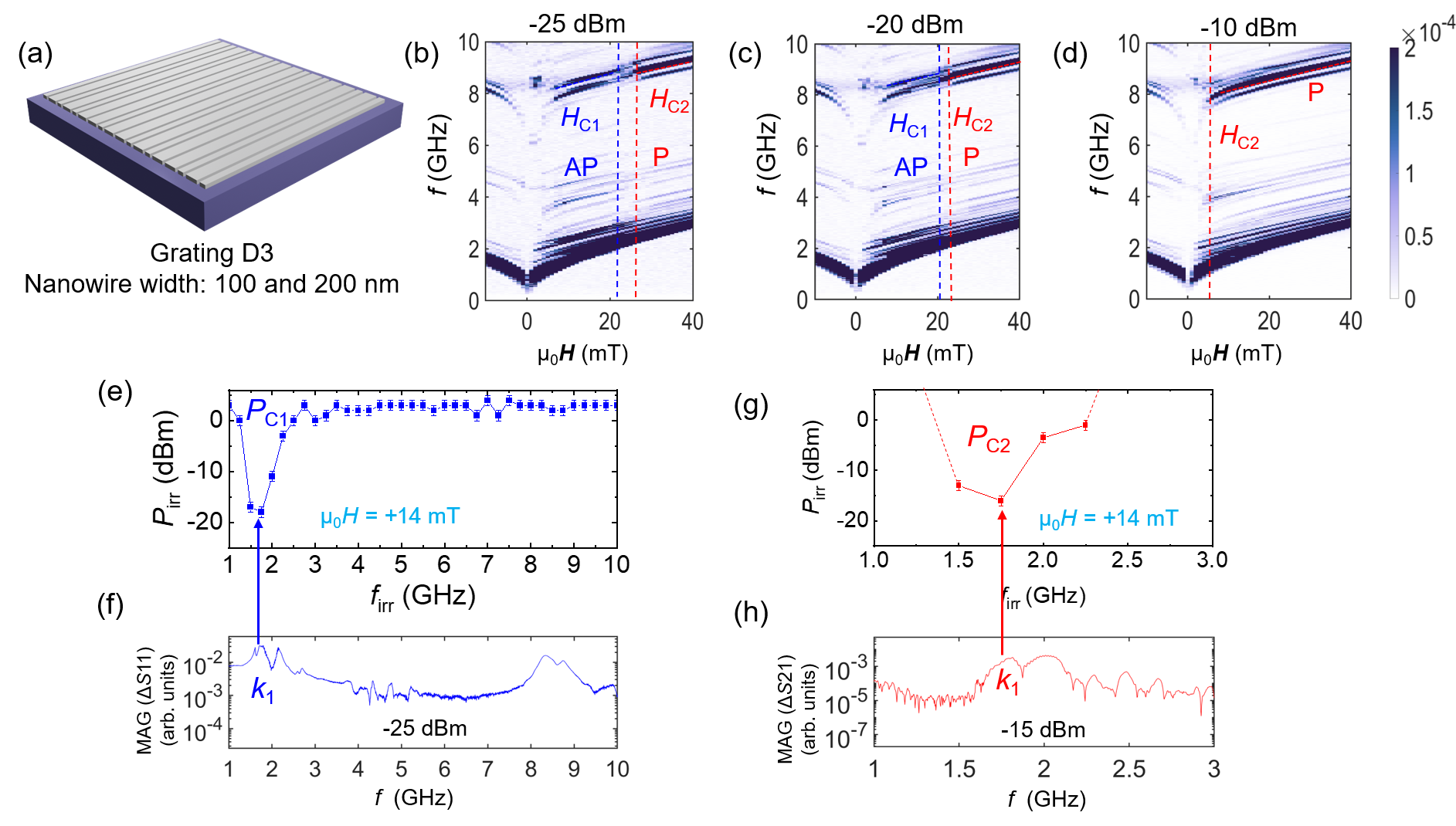}
    \caption{(a) Schematic of sample D3 with a Fibonacci grating consisting 100 nm and 200 nm wide Py nanowires separated by 100 nm, (b-d) plots of magnitude of $\Delta S21$ spectra measured for the device at increasing applied microwave powers ($P_{\rm irr}$), (e) Switching yield diagram (in blue) showing a trend of the critical power that corresponds to switching of the gratings beneath CPW1 by $k_1$ mode (f) Magnitude of the reflection spectrum $\Delta S11$ measured at +14 mT showing the $k_1$ mode (g) Switching yield diagram (in red) of the grating beneath CPW2 which corresponds to switching by $k_1$ mode that propagates from CPW1 to CPW2 as observed from (h) magnitude of $\Delta S21$ measured at +14 mT. Dotted lines in (g) are a guide to the eye. They indicate that the critical power $P_{\rm C2}$ for the modes excited at $f_{\rm irr}<1.5~$GHz and $>2.5~$GHz is beyond +6~dBm}
    \label{fig:fig3}
\end{figure}

In sample D3 nanostripes were arranged in a quasi-crystalline Fibonacci sequence beneath both the CPWs [Fig. \ref{fig:fig3}(a)]. The nanostripes were 100 nm and 200 nm wide separated by a 100-nm-wide air gap.
Figure \ref{fig:fig3} (b) shows the magnitude of $\Delta S21$ between 10 MHz to 10 GHz measured at $P_{\rm irr}=-25$~dBm. Dark branches correspond to the spin wave modes propagating in YIG from CPW1 to CPW2. Multiple spin wave modes are resolved. These modes are excited due to the CPW (mode $k_1$) as well as reciprocal vectors \cite{SWatanabe2023} provided by the Fibonacci grating beneath CPW. Again, the high-frequency modes can be divided into low-field AP branches and high-field P branches. Similar to Fig. \ref{fig:fig1}(d-e), branches vanish and reappear when the applied magnetic field is swept from $-30$~mT to $+40$~mT. Branches chosen for the analysis of the nanowires' switching fields are marked by a dashed blue line (AP mode) for the onset of switching and by a dashed red line (P mode) for the completion of switching. Figure \ref{fig:fig3}(b-d) shows spectra for step-wise increased microwave power $P_{\rm irr}$. $H_{\rm C1}$ (blue dashed line) and $H_{\rm C2}$ (red dashed line) decrease with $P_{\rm irr}$. The AP branch approaches zero field for $P_{\rm irr}$ above -15 dBm, i.e., at a smaller power compared to D1, but similar to D2 (see below). We attribute this observation to a smaller reversal field of 200-nm-wide nanostripes compared to the narrower ones in the aperiodic array.

Switching yield experiments were performed on D3 similar to D1 but at an opposing field of +14 mT. The sample D3 was saturated at $\mu_0H$= -90 mT and the field was swept to +14 mT, when the grating was in AP configuration. We used intentionally the same field value like in Ref. \cite{Baumgaertl2023} to compare directly with the reversal reported for a periodic grating of 100-nm-wide nanostripes. Then, the spin waves were excited at irradiation frequency $f_{\rm irr}$ ranging from 1 GHz to 10 GHz with a 0.25 GHz step. At each $f_{\rm irr}$, the microwave power $P_{\rm irr}$ was increased from -25 dBm to +6 dBm. Switching events were recorded by measuring the magnitude of $\Delta S21$ in the frequency window of 2.5 to 5.5 GHz. The critical powers at which spin wave signals first reduced and then increased to 50\% of the maximum signal strength were denoted as $P_{\rm C1}$ and $P_{\rm C2}$, respectively. The symbols displayed as a function of $f_{\rm irr}$ in Fig. \ref{fig:fig3}(e) [\ref{fig:fig3}(g)] reflect the onset (completion) of reversal of stripes beneath CPW1 (CPW2). The switching yield diagrams show that mode $k_1$ induces reversal of nanostripes below CPW1 at 15.8~$\mu$W and below CPW2 at 25.1 $\mu$W. The latter value is a factor of two smaller compared to Ref. \cite{Baumgaertl2023} where periodically arranged stripes with $w=100~$nm were reversed underneath CPW2 by mode $k_1$ in a field of $+14~$mT. Grating coupler modes in D3 did not allow us to reverse nanostripes beneath CPW2 up to the value of 50\% of the branch's signal strength. This is unlike the observation in case of sample D1 with periodically arranged 50-nm-wide nanowires [see Fig. \ref{fig:fig2}(c)] where a grating coupler mode switched the remotely placed stripes at $P_{\rm C2}$.

\subsection{Magnon-induced switching dependent on nanostripe width}

\begin{figure}[H]
    \centering
    \includegraphics[width= \linewidth]{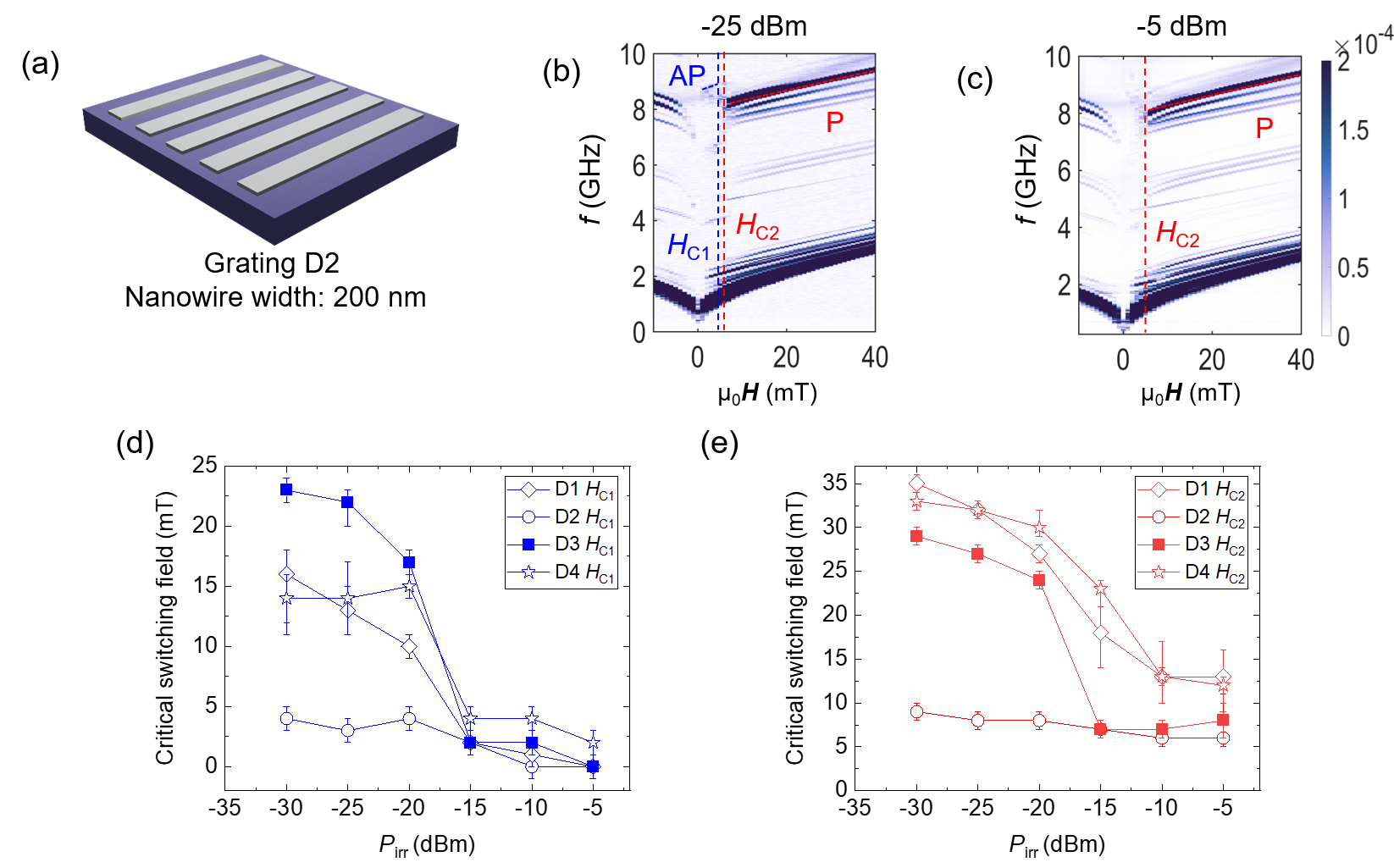}
    \caption{(a) Sketch of sample D2 with periodic gratings consisting of 200-nm-wide Py nanostripes arranged with $a=400$ nm. $\Delta S21$ spectra measured on D2 at an applied power $P_{\rm irr}$ of (b) -25 dBm and (c) -5 dBm. (d) Critical field $H_{\rm C1}$ and (e) $H_{\rm C2}$ extracted for nanostripes with widths of 50 nm (D1), 100 nm (D4 adopted from  Ref \cite{Baumgaertl2023}), and 200 nm (D2) denoted by hollow symbols as well as the aperiodic grating (D3) denoted by solid rectangle.}
    \label{fig:fig4}
\end{figure}

In this section, we compare the critical switching fields for samples incorporating different nanostripe widths. Figure \ref{fig:fig4}(a) to (c) show the parameters and spin-wave spectra obaianed on sample D2 with $a=200$ nm and $w=400~$nm. The switching fields $H_{\rm C1}$ and $H_{\rm C2}$ reduce with power $P_{\rm irr}$. However, their values and variation are much smaller than for D1. In Fig. \ref{fig:fig4}(d) and (e) we summarize power-dependent critical fields extracted for three different samples D1, D2 (hollow symbols), D3 (solid rectangle) and compare them to sample D4 reported in Ref. \cite{Baumgaertl2023} (star). A decreasing trend in $H_{\rm C1}$ and $H_{\rm C2}$ with $P_{\rm irr}$ is observed for all samples substantiating magnon-induced reversal for nanostripes of different widths between 50 and 200 nm. 
A detailed analysis provides two key observations:
\begin{enumerate}
    \item D1, D3 and D4 show a sharp reduction in $H_{\rm C1}$ and $H_{\rm C2}$ as compared to D2. The net reduction in $H_{\rm C2}$ due to spin wave excitation between -25 dBm to -5 dBm in D1, D3 and D4 is about five times larger than in D2 (see supporting information S3). 
    \item In Fig. \ref{fig:fig4}(d) and (e), sample D1 with $w=50~$nm behaves similarly to D4 with $w=100~$nm presented in Ref. \cite{Baumgaertl2023} (star). Only D2 with $w=200~$nm shows critical fields which are much lower at low power than all other samples. Topp et al. in Ref. \cite{Topp2009Neff} presented an analytical model allowing us to calculate the effective transverse demagnetization factors $N_{\rm eff}$ for nanostripes arranged in periodic lattices. The model quantifies the effect of dipolar interaction which is known to modify magnetic ansiotropies \cite{PhysRevB.63.104415}. We find 0.2 for D1, 0.1 for D4 and 0.05 for D2. Assuming coherent rotation of spins, the magnetic anisotropy fields should scale accordingly \cite{Adeyeye96,Tannous_2008_Stonermodel}, suggesting a clearly larger reversal field for D1 compared to D4 and D2. We find however, that samples D1 and D4 show very similar critical fields $H_{\rm C2}$ at the lowest VNA power. This observation suggests that the reversal mechanism involves incoherent processes (nonuniform rotation) like spin curling \cite{aharoni} or domain wall motion \cite{Hayashi2007DW,Fan2023_DW}.
\end{enumerate}

\noindent Figure \ref{fig:fig4} shows that magnon-induced reversal occurred for both the narrowest nanostripe with a large reversal field of up to about 35 mT and the wide ones with a reversal field of 10 mT and smaller. The absolute reversal fields of the different samples do not scale according to their nanostripe widths and effective demagnetization factors indicating a reversal mechanism involving incoherent spin rotation. The microscopic mechanism and relevant torque leading most efficiently to magnon-induced reversal of nanomagnets are still under debate and not decided by the experiments presented here. Importantly, our experiments showed that one and the same 50-nm-wide Py nanostripes were reversed by propagating spin waves in YIG which exhibited wavelengths $\lambda$ varying by a factor of 70. In case of $\lambda\approx 100$~nm, the propagation distance was $\geq 250\times \lambda$. Such a macroscopically large distance allows for both coherent scattering in a nanostructured holographic memory \cite{khitun2023} and computation across several unit cells of a neural network \cite{csaba} before inducing nonvolatile storage of the computational result in a separate magnetic bit.   

\subsection{Conclusion}
Power dependent spectroscopy performed on thin YIG containing arrays of Py nanostripes with widths $w$ ranging from 50 to 200 nm evidenced their magnon-induced reversal in small opposing fields. We observed a sharp decrease in the reversal fields of stripes with $w=50~$nm after applying a VNA power of about -20 dBm (10 $\mu$W) to the spin-wave emitting coplanar waveguide. The sharp decrease in switching field was observed also in the sample with an aperiodic arrangement of nanostripes with $w=100~$nm and 200~nm. On the other hand, a more steady decrease of reversal field was seen in case of periodic arrays with 200-nm-wide nanowires. However, their initial reversal field was already small. Spin waves excited directly by the CPW exhibited $\lambda\approx 7~\mu$m and reversed nanostripes in all the four investigated samples. Magnons with $\lambda\approx 100$~nm induced efficiently the reversal of 50-nm-wide Py stripes. The required precessional power was as low as 10 nW in a small opposing field of +10 mT. We highlight that switching of all the different Py nanostripes was achieved by spin waves which propagated over 25 $\mu$m in YIG away from the spin wave emitter. Our findings are important for nanoscale magnonic in-memory computation, the holographic memory and reconfigurable spin wave-based computing devices.


\newpage

\section{Methods}
\subsection{Sample fabrication}
We used a 100-nm-thick YIG thin film epitaxially grown on a GGG (111) substrate. It was provided by the Matesy GmbH in Jena, Germany. A 20-nm-thick Py (Ni$_{81}$Fe$_{19}$) film was deposited on the YIG film using electron beam evaporation. Two groups of gratings of nanostripes separated by 25 $\mu$m were etched using a resist mask prepared via electron beam lithography applied to hydrogen silsesquioxane. The sample D1 (D2) was patterned with 50 nm (200 nm) wide nanostripes periodically arranged in a 100 nm (400 nm) period. The lengths of nanostripes were alternately varied between 25 and 27 $\mu$m. D3 was fabricated with 25 $\mu$m long 100 nm and 200 nm-wide nanostripes arranged in Fibonacci sequence with a gap of 100 nm. The samples were etched with an Ar ion beam considering an etch stop at the YIG film.  Two Coplanar Waveguides (CPWs) were patterned above the nanostripe gratings with signal-to-signal line separation of 35 $\mu$m using electron beam lithography with an MMA-PMMA double layer positive resist, electron beam evaporation of 5 nm thick Ti and 110 nm thick Cu followed by lift-off processing. Ti was deposited for adhesion. 

\subsection{Broadband Spin Wave Spectroscopy}

The magnon-induced reversal was investigated using Keysight Vector Network Analyzers (VNAs) N5222A and N5222B. The probe station of the VNA setup consisted of electromagnetic pole shoes which provided a bipolar magnetic field of up to 90 mT at arbitrary in-plane directions. The CPW1 or emitter CPW of the sample was connected to the VNA port 1 with a microprobe. CPW2 is connected to the detector port 2 with an identical microprobe. Initially, the sample was saturated at -90 mT (in $-y$ direction) and spin waves were excited with microwaves of frequencies ranging from 10 MHz to 20 GHz with a step of 2.5 MHz. The scattering parameters $S11$ and $S21$ in reflection and transmission configuration, respectively, were recorded. The static field was brought to -30 mT and increased to +40 mT in a step-wise manner. At each step, the scattering parameters were recorded. The measurements were carried out for spin wave modes excited with applied microwave powers from -25 dBm to +6 dBm. These all-electrical spin-wave spectroscopy (AESWS) measurements were repeated for devices with different grating periods and nanostripe widths and also for gratings with nanowires arranged in Fibonacci sequence. After every measurement, the median subtracted magnitude of scattering parameters as a function of static magnetic field were plotted and analyzed for different spin wave modes. The critical switching fields, i.e. static magnetic fields at which the gratings started ($H_{\rm C1}$) and completed reversal ($H_{\rm C2}$)  were decided based on the disappearance and reappearance of a specific spin wave branch for all devices between 8-10 GHz at magnetic fields swept up to $+40~$mT. 

\subsection{Switching yield diagram}
We performed switching yield diagram measurements as follows. A positive opposing field $\mu_0H$ was applied to a sample after saturation at negative fields. 
A transmission spectrum $S21$ was measured within a frequency window of 6-9.5 GHz for D1 and 2.5-6.5 GHz for D3 ($f_{\rm sens}$) excited at -25 dBm ($P_{\rm sens}$). We call this operation as sensing of the nanostripes' magnetic configuration being either anti-parallel (before) or parallel to the small bias field (after magnon-induced reversal). Spin waves were excited with irradiation frequency ($f_{\rm irr}$) range of 1-1.25 GHz at an irradiation power of -25 dBm ($P_{\rm irr}$). After the irradiation, a  $S{21}$ transmission spectrum was measured in the sensing window ($f_{\rm sens}$)  to sense the effect of the irradiation on the nanostripes. This measurement sequence was repeated for the spin wave mode excited at an increasing $P_{\rm irr}$ from -25 dBm to the maximum of +6 dBm in case of the D3 and up to +15 dBm in case of the D1. The magnetic field protocol was reset and the measurements were repeated at the next interval of irradiation frequency ($f_{\rm irr}$) up to 10 GHz. As an example, median subtracted magnitude $\Delta S21$ transmission spectra plotted for D1 measured at $f_{\rm irr}$ from 1.75- 2 GHz across the range of $P_{\rm irr}$ are provided in the supporting information S1. The spin wave irradiation powers inducing a disappearance and reappearance of spin wave branches in these plots corresponded to critical powers needed to start ($P_{\rm C1}$) and complete ($P_{\rm C2}$), respectively, the switching of nanostripes below CPW1 and CPW2, respectively. 

\begin{acknowledgement}

The authors thank for support provided by the Centre of MicroNanoTechnology (CMi) during the fabrication of the devices. Authors thank Mingran Xu and Mohammad Hamdi for valuable discussions. The AESWS spectra were visualized using colourmaps provided by Fabio Crameri \cite{Crameri_2023}. The scientific colour map devon is used to prevent visual distortion of the data. The research was supported by the SNSF via grant number 197360.
\section{Data availability statement}

The data that support the findings of this study are available from the corresponding author upon reasonable request. 


\section{Conflict of interest disclosure}
The authors declare no conflict of interest.

\subsection*{Author contributions}
S.J., K.B., and D.G. planned the experiments. K.B. and D.G. designed the samples. K.B. prepared the samples. S.J. performed the experiments together with K.B., F.B. and A.M. S.J., F.B., A.M. and D.G. analyzed and interpreted the data. S.J. and D.G. wrote the manuscript. All authors commented on the manuscript.



\end{acknowledgement}

\newpage 
\begin{suppinfo}
\subsubsection{S1: Criteria for extraction of critical powers $P_{\rm C1}$ and $P_{\rm C2}$ from switching yield diagram experiment for device D1}
The switching yield diagram is a plot of critical power needed to start and complete the switching of gratings beneath both the CPWs. This is calculated with the following strategy: 
\begin{itemize}
    \item The device is saturated at -90 mT and applied magnetic field is swept up to +10 mT (for D1). Median subtracted magnitude of $\Delta S21$ is measured at $P_{\rm sens} =-25~$ dBm and in certain frequency window ($f_{\rm sens}$: 6 to 9.5 GHz), where the magnon modes are detected [black branches in Fig. \ref{fig:si1}(a)]. Then, a specific magnon mode between certain frequency window $f_{\rm irr}$ (here, 1.75 to 2 GHz) and for $P_{\rm irr}$ from $-25~$dBm to $+15~$dBm is irradiated. After every irradiation step at $P_{\rm irr}$, magnitude of $\Delta S21$ is measured at $P_{\rm sens}=-25~$ dBm and frequency window $f_{\rm sens}$.
    \item The criteria of extracting $P_{\rm C1}$ and $P_{\rm C2}$ is explained based on the disappearance of magnon modes and reappearance at a new frequency highlighted with blue and yellow brackets respectively.
\end{itemize} 
\begin{figure}[H]
    \centering
    \includegraphics[width= \linewidth]{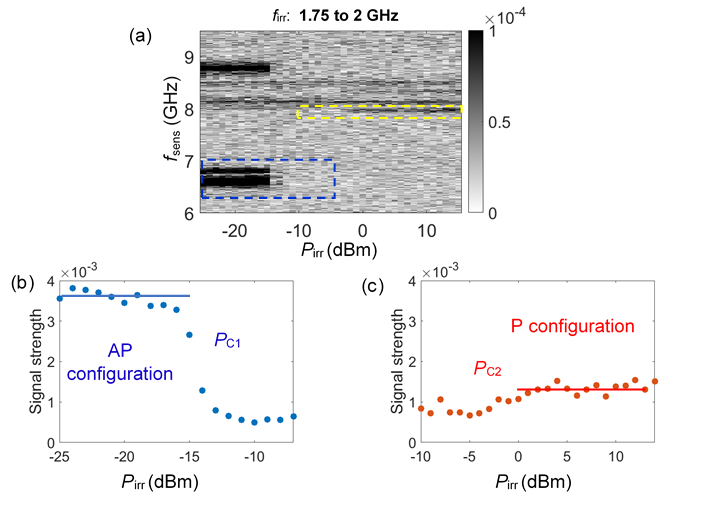}
    \caption{Procedure for determining the critical switching powers ($P_{\rm C1}$ and $P_{\rm C2}$) from the $S$21 transmission measurement after exciting a magnon mode from 1.75 to 2 GHz at an increasing irradiation power ($P_{\rm irr}$). The two black branches at between 6.5-7 GHz and 8.5-9 GHz correspond to the spin wave modes sensed for $P_{\rm irr}$ below $-15~$dBm. The integrated signal strength from $f_{\rm sens}$ between 6.4 to 7 GHz (marked with blue dotted lines) was extracted and plotted as a function of $P_{\rm irr}$ in Fig. (b) showing the disappearance of the mode between $P_{\rm irr}:-25$ to $-10~$ dBm. $P_{\rm C1}$ corresponds to $P_{\rm irr}$ at which the mode is at the average of its maximum signal strength and minimum signal strength or noise floor. For the given $\Delta S21$ spectra, it is $-14~$dBm with an error bar of $\pm 1~$ dBm. On the contrary, a spin wave mode appears at 8 GHz for $P_{\rm irr}>0~$dBm. We apply similar criteria but now for the increase in the signal strength for the branch highlighted in yellow dotted lines. The signal strength was extracted between 7.8 to 8.04 GHz and plotted in Fig. (c) as a function of $P_{\rm irr}$. The critical switching power at which a new mode appears is denoted as $P_{\rm C2}$ and is given by $50\%$ of maximum signal strength of the branch which corresponds to 0~dBm with an error bar of $\pm 1~$ dBm.}
    \label{fig:si1}
\end{figure}

\subsubsection{S2: Calculation of precessional power ($P_{\rm prec}$) at $P_{\rm C1}$ and $P_{\rm C2}$ for device D1}
Precessional power for a magnon mode with frequency $f_{\rm irr}$ excited at microwave power ($P_{\rm irr}$) is given by: 

\[P_{\rm prec} = (\rm MAG(\Delta\textit{S}11)~\rm at~\textit{f}_{\rm irr})^2 \times (\textit{P}_{\rm C}~at~ \textit{P}_{\rm irr});\]

\noindent Where, $P_{\rm C1}$ and $P_{\rm C2}$ are obtained from switching yield diagram measurement as shown in the previous section. To calculate magnitude of $\Delta S$11 [MAG ($\Delta S11$)], we subtract the RAW data using reference subtraction from the spectra measured at zero field. The following plot shows reference subtracted $\Delta S11$ spectra.

\begin{figure}[H]
    \centering
    \includegraphics[scale= 1.0]{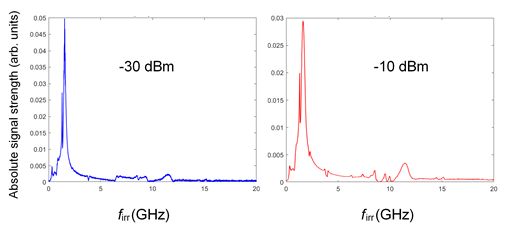}
    \caption{Magnitude of $\Delta S11$ spectra at 10 mT measured at $-30~$dBm and $-10~$dBm after reference subtraction from measurement 0 mT. The signal strength for both $k_1$ and $|G^{\rm D1} - k_1|$ modes were extracted for evaluating $P_{\rm prec,1}$ and $P_{\rm prec,2}$}
    \label{fig:si2}
\end{figure}

\noindent $P_{\rm prec,1}$ was calculated with $\Delta S11$ spectra measured at $-30~$dBm, where, both the gratings are antiparallel to YIG and the applied field direction. Whereas, for $P_{\rm prec,2}$, we use the spectra measured at $-10~$dBm. At this microwave power the gratings are completely switched beneath both the CPWs.

\subsubsection{S3: Extent of critical field reduction extracted from $\Delta S21$ spectra of devices D1 (consisting 50 nm wide nanostripes) and D2 (consisting 200 nm wide nanostripes) measured at different microwave powers $P_{\rm irr}$}
\begin{figure}[H]
    \centering
    \includegraphics[scale= 1.0]{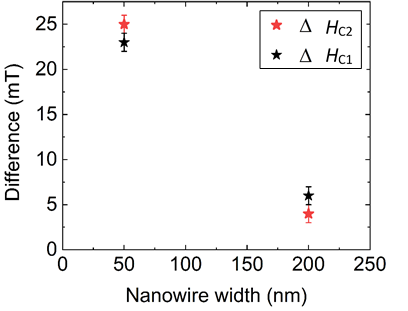}
    \caption{The trend of the difference in critical switching fields extracted for the devices D1 and D2 consisting of 50 nm wide and 200 nm wide Py nanostripes respectively. The data points are calculated from the $H_{\rm C1}$ and $H_{\rm C2}$ values obtained from $\Delta S21$ spectra. The trend of this values is shown in the main text in Fig. \ref{fig:fig4}(d-e).}
    \label{fig:si3}
\end{figure}

\end{suppinfo}
\newpage

\bibliography{achemso-demo}

\end{document}